\documentclass{article}

\usepackage{arxiv}

\usepackage[utf8]{inputenc} % allow utf-8 input
\usepackage[T1]{fontenc}    % use 8-bit T1 fonts
\usepackage{hyperref}       % hyperlinks
\usepackage{url}            % simple URL typesetting
\usepackage{booktabs}       % professional-quality tables
\usepackage{amsfonts}       % blackboard math symbols
\usepackage{nicefrac}       % compact symbols for 1/2, etc.
\usepackage{microtype}      % microtypography
\usepackage{graphicx}
\usepackage{natbib}
\usepackage{doi}
\usepackage{balance} % for balancing columns on the final page
\usepackage{multirow}
\usepackage{amsthm}
\usepackage{amsmath}
\usepackage{enumitem}
\usepackage{xcolor}
\usepackage[english]{babel}

\newtheorem{theorem}{Theorem}[section]

\newtheorem{corollary}[theorem]{Corollary}

\newenvironment{definition}[1][Definition]{\begin{trivlist}
\item[\hskip \labelsep {\bfseries #1}]}{\end{trivlist}}
\newenvironment{remark}[1][Remark]{\begin{trivlist}
\item[\hskip \labelsep {\bfseries #1}]}{\end{trivlist}}

\title{Multi-unit Double Auctions: Equilibrium Analysis and Bidding Strategy using DDPG in Smart-grids}

\author{{Sanjay Chandlekar}\\
	Machine Learning Lab, IIIT Hyderabad \\
	Hyderabad, India \\
	\texttt{sanjay.chandlekar@research.iiit.ac.in} \\
	\And
	{Easwar Subramanian} \\
	TCS Innovation Labs\\
	Hyderabad, India \\
	\texttt{easwar.subramanian@tcs.com} \\
	\AND
	{Sanjay Bhat} \\
	TCS Innovation Labs \\
	Hyderabad, India \\
	\texttt{sanjay.bhat@tcs.com} \\
	\And
	{Praveen Paruchuri} \\
	Machine Learning Lab, IIIT Hyderabad \\
	Hyderabad, India \\
	\texttt{praveen.p@iiit.ac.in} \\
	\And
	{Sujit Gujar} \\
	Machine Learning Lab, IIIT Hyderabad \\
	Hyderabad, India \\
	\texttt{sujit.gujar@iiit.ac.in} \\
}

\date{}

\renewcommand{\shorttitle}{\raggedright Sanjay Chandlekar et al.}
\newcommand{\ddpg}{\textsl{DDPGBBS}}
\newcommand\numberthis{\addtocounter{equation}{1}\tag{\theequation}}

\hypersetup{
pdftitle={Multi-unit Double Auctions: Equilibrium Analysis and Bidding Strategy using DDPG in Smart-grids},
pdfsubject={cs.GT, econ.TH},
pdfauthor={Sanjay ~Chandlekar, Easwar ~Subramanian, Sanjay ~Bhat, Praveen ~Paruchuri, Sujit ~Gujar},
pdfkeywords={Multi-unit Periodic Double Auction, Equilibrium Analysis, Bidding Strategy, DDPG for Learning Equilibrium Strategy},
}

\begin{document}
\maketitle

\begin{abstract}
	Periodic double auctions (PDA) have applications in many areas such as in e-commerce, intra-day equity markets, and day-ahead energy markets in \emph{smart-grids}. While the trades accomplished using PDAs are worth trillions of dollars, finding a reliable bidding strategy in such auctions is still a challenge as it requires the consideration of future auctions. A participating buyer in a PDA has to design its bidding strategy by planning for current and future auctions. Many equilibrium-based bidding strategies proposed are complex to use in real-time. In the current exposition, we propose a scale-based bidding strategy for buyers participating in PDA. We first present an equilibrium analysis for single-buyer single-seller multi-unit single-shot k-Double auctions. Specifically, we analyze the situation when a seller and a buyer trade two identical units of quantity in a double auction where both the buyer and the seller deploy a simple, scale-based bidding strategy. The equilibrium analysis becomes intractable as the number of participants increases. To be useful in more complex settings such as wholesale markets in smart-grids,  we model equilibrium bidding strategy as a learning problem. We develop a \emph{deep deterministic policy gradient (DDPG)} based learning strategy, \ddpg, for a participating agent in PDAs to suggest an action at any auction instance. \ddpg, which empirically follows the obtained theoretical equilibrium, is easily extendable when the number of buyers/sellers increases. We take Power Trading Agent Competition's (PowerTAC) wholesale market PDA as a testbed to evaluate our novel bidding strategy. We benchmark our DDPG based strategy against several baselines and state-of-the-art bidding strategies of the PowerTAC wholesale market PDA and demonstrate the efficacy of \ddpg\ against several benchmarked strategies. 
\end{abstract}

% keywords can be removed

\section{Introduction}
    
    \hspace{0.5cm}An auction is a process of buying or selling goods or items~\cite{auction_wiki}, and it could be of varied types~\cite{narahari_auction_types}. For example, a \emph{forward auction} consists of one seller and multiple buyers who submit their \emph{bids}; a \emph{reverse auction} has one buyer and multiple sellers submitting their \emph{asks};  a \emph{double auction} comprising of multiple buyers and sellers placing their bids and asks. Double auctions are extensively used in the real world to trade stocks, energy, spectrum, and many other goods and services. Stock Exchanges are one example where a double auction generally takes place and involves trillions of dollars daily trades. The other prominent domain where the double auction plays a significant role is \emph{smart-grids}~\cite{smartgrid}. 
    
    \hspace{0.5cm}Multiple power generating companies (GenCos) and energy brokers trade energy in the wholesale market through double auction in smart grids. Nord Pool, which runs the leading power market in Europe, showed trades of $995$ $TWh$ of volume, with close to 60\% of the volume traded using APIs ~\cite{nordpool}. Clearly, a bidding strategy that can optimize the cost of energy brokers even by a small amount would significantly improve their profits and bring in more efficiency to the system. While such double auctions are common in the real world, designing an optimal bidding strategy that an energy broker (buyer) can use in real-time applications is still challenging. The problem gets further complicated in settings where periodic double auctions are deployed. In  PDAs, the buyer needs to participate in a series of auctions, and therefore, a  bidding strategy involves planning across current and future auctions. For example, an energy broker participating in day-ahead markets can procure energy for a delivery time-slot at any prior time within 24 hours. 
    
    \hspace{0.5cm}Motivated with such smart-grid energy trading, we focus on scaling-based bidding strategies for the following reasons: -  (i) they are the most natural when the brokers need to submit bids in real-time. (ii) as compared to solving differential equations~\cite{vetsikas} or fictitious play formulations~\cite{soton270991} to devise strategies, which are complex in nature, scale-based strategies are easy to implement. Hence, the primary reason to use scale-based strategies is their simplicity as an agent needs to bid in real-time in settings like smart-grids. Ghosh et al.~\cite{susobhan20} already designed a scale-based bidding strategy for single unit double auction and claim to have a good performance, making them runner-up in the PowerTAC 2018 tournament. We study a two-unit double auction with a single seller and a single buyer deploying scaling-based strategies and characterize Nash equilibria (NE) for the buyer and the seller when they trade two units ($m=2$) of identical and indivisible items at each instance of a $k$-double auction. Characterizing equilibrium becomes intractable beyond $m=2$. Hence, we address the issue of whether an intelligent agent (buyer) can learn the equilibrium strategy.
    
    \hspace{0.5cm}Motivated by the recent success stories of employing neural networks (NN) to solve game theoretical problems \cite{manisha1,manisha2,manisha3,parks}, we use techniques from \emph{reinforcement learning} and, more specifically, we develop a deep deterministic policy gradient-based bidding strategy to address the issue of learning equilibrium strategy. The strategy thus developed is then tested on a smart-grid wholesale market ecosystem called PowerTAC, which is an efficient implementation of the real-world smart-grid~\cite{KETTER2013262}. PowerTAC replicates the crucial elements of smart-grids; more importantly, it has PDA for energy trading in the wholesale energy market. Past works that were used in the PowerTAC framework include MDP-based bidding strategies used in TacTex~\cite{tactex2013} and VidyutVanika (called MDPLCPBS)~\cite{susobhan20}, an MCTS based formulation in SPOT~\cite{ijcai2018-23} among others. These prior strategies discretize the action space, which is typically continuous in nature. In addition, except SPOT, most other works treat the bidding process as a single unit auction. Whereas, in this work, we consider the setting of a two-unit auction (dividing the required quantity into two equal parts) as we believe it helps in bringing down the clearing price of an auction. %better to leverage similar to a volume discount \ea{Not clear about this statement}. 
    With this background, we design a Deep-Deterministic-Policy-Gradient (DDPG) based bidding strategy -- \ddpg\  for double auctions. We chose DDPG, as it has proved to be very successful in a variety of \emph{planning} tasks~\cite{lillicrap2016}. \ddpg\ learns to bid optimally with the help of feedback from the auction environment. 
    
    \hspace{0.5cm}To test the efficacy of \ddpg\ to learn equilibrium strategies, first, we conduct validation experiments to confirm if it learns the known equilibria. We observe that it can approximate the NE under the controlled experimental setting for all the known cases. We designed a simulator comprising one buyer, one seller, and a two-unit double auction to perform validation experiments. We model the seller to follow its NE strategy and train the agent (buyer) against it. 
    
    \hspace{0.5cm}Post verifying the ability to learn equilibrium bidding strategy with \ddpg; we extend it to model general PDA where there is no restriction on the number of participants or the number of units traded in the auction. As there are multiple opportunities for the buyer to buy energy for a particular delivery time-slot, we introduce \emph{proximity} to the delivery time-slot into our state-space formulation. For training and validating \ddpg\ for such real-world scenarios, we leverage the PowerTAC PDA simulation environment to conduct all our experiments instead of building our simulator. Many researchers have attempted to design a bidding strategy for the PowerTAC PDA; previous bidding strategies like MCTS based strategy of SPOT~\cite{ijcai2018-23}, MDPLCPBS of VidyutVanika~\cite{susobhan20} and MDP based strategy of TacTex~\cite{tactex2013} are proven to be effective in the annual PowerTAC tournament. However, they do not perform game-theoretical equilibrium analysis for their strategies except VidyutVanika. Although VidyutVanika demonstrated that their MDPLCPBS follows theoretical NE, their work was limited to single-unit demand. 
    
    \hspace{0.5cm}To the best of our knowledge, we are the first to utilize the policy gradient-based RL algorithm, enabling us to work with PowerTAC's continuous state and action space more effectively. In summary, our contributions are as follows:
    \begin{enumerate}
        \item We analyze Bayesian Nash equilibrium strategies for one buyer and one seller in multi-unit auctions, where both buyer and seller trade for two identical items by practicing scale-based bidding strategies. We present an analysis for each of the four possible scenarios depending on whether the buyer and seller's scale factors are the same for both units or different.  
        \item We formalize a DDPG based bidding strategy (\ddpg) for multi-unit double auction and validate that it learns the theoretical equilibrium.
        \item We extend \ddpg\ to work effectively in PDAs such as the PowerTAC wholesale market. We then demonstrate its efficacy against baseline and state-of-the-art bidding strategies for PowerTAC.  
    \end{enumerate}

\section{Related Work}

   \hspace{0.5cm}Chatterjee and Samuelson~\cite{chatterjee1983} propose a Nash Equilibrium analysis for a single buyer and single seller with the general distribution of valuations and constructed NE strategy for a special case of uniform valuations. Satterthwaite and Williams~\cite{satterthwaite1989} address the problem of finding the existence of a multiplicity of equilibria for the $k$-double auction for a generic class of traders' valuations; they too propose the equilibrium strategies in the form of differential equations and then examine the efficiency of the proven equilibrium. Vetsikas~\cite{vetsikas2014} attempts to find equilibrium strategies for multi-unit sealed bid auction for $m^{th}$ and $(m+1)^{th}$ price sealed bid auction. 
    
   \hspace{0.5cm}As for bidding strategies in PowerTAC PDA, Kaute et al.~\cite{rodrigue2013} uses an MDP-based bidding strategy in their broker AstonTAC, which outputs the ratio of remaining energy that needs to be ordered, while limit-price comes from the  Non-Homogeneous Hidden Markov Models (NHHMM). Urieli and Stone~\cite{tactex2013} propose an MDP-based strategy inspired by Tesauro and Bredin's bidding strategy~\cite{tesauro2002} for broker TacTex, and they solve this MDP using dynamic programming. Ghosh et al.~\cite{susobhan20,susobhan2} improve upon TacTex's bidding strategy to use in the VidyutVanika broker; additionally, they provide theoretical analysis for their MDP-based strategy. Chowdhury et al.~\cite{spot18}~\cite{spot_aamas} use a Monte Carlo Tree Search (MCTS) based strategy on top of REPTree based price predictor coupled with heuristics to determine the optimal bid prices. Ozdemir and Unland~\cite{agentude2015} use an adaptive Q-learning-based strategy in their broker AgentUDE. TUC\_TAC~\cite{tuctac2021} too uses an MCTS based bidding strategy inspired by Chowdhury et al. in the wholesale market. 
    
   \hspace{0.5cm}The existing theoretical equilibrium for multi-unit double auction are complex and may not be readily helpful in real-world double auctions. In contrast, our scale-based bidding strategy is easy to implement for such auctions. Furthermore, none of the previous works attempt to design a bidding strategy for multi-unit double auctions backed up by game-theoretic analysis; we show that our novel \ddpg\  based algorithm works effectively in double auctions while closely following the theoretical equilibria.
    
%%%%%%%%%%%%%%%%%%%%%%%%%%%%%%%%%%%%%%%%%%%%%%%%%%%%%%%%%%%%%%%%%%%%%%%%

\section{Preliminaries}

    Here, we define all the terminologies and notations used in this paper.
    
    %%%%%%%%%%%%%%%%%%%%%%%%%%%%%%%%%%%%%%%%%%%%%%%%%%%%%%%%%%%%%%%%%%%%%%%%

    \subsection{k-Double Auction}
        \hspace{0.5cm}k-double auction is a type of auction for buying and selling resources, where potential buyers submit their bids and potential sellers submit their asks to the auctioneer. The auctioneer receives all the bids and asks, and determines each player's \emph{clearing quantity} and the \emph{clearing price} using a specific allocation rule and payment rule. 
        \begin{definition}[Clearing Price]
            A clearing price is a price at which the auctioneer clears the market after matching potential bids of the buyers with the potential asks of the seller.
        \end{definition}
        \begin{definition}[Clearing Quantity]
            A clearing quantity of a player is the number of items a buyer[seller] receives[sells] from the auction after clearing.
        \end{definition}
        \begin{definition}[Allocation Rule]
            An allocation rule determines the quantity bought by each buyer and the quantity sold by each seller in an auction. Basically, the allocation rule determines the clearing quantity of each player.
        \end{definition}
        \begin{definition}[Payment Rule]
            A payment rule determines the payment each player has to make (buyer pays and seller earns) at the time of auction clearing. 
        \end{definition}
        \begin{definition}[$k$-Double Auction]
            If a buyer and seller participate in a double auction, and if the buyer's bid $b$ is higher than the seller's bid $s$, then the clearing price is given by $kb+(1-k)s$ for some fixed $k \in [0, 1]$.
        \end{definition}

        \hspace{0.5cm}In this paper, we consider a specific case of \emph{k-double auction} for \emph{$k$=0.5}, which we refer to as the \textbf{average clearing price rule (ACPR)}. This implies that the auction clearing price for $k=0.5$ would be an average of the last clearing bid and the last clearing ask. Additionally, we assume the uniform pricing rule, which says that all the buyers[sellers] pay[receive] the same price for a unit quantity. 

    %%%%%%%%%%%%%%%%%%%%%%%%%%%%%%%%%%%%%%%%%%%%%%%%%%%%%%%%%%%%%%%%%%%%%%%%

    \subsection{Notation}
        \hspace{0.5cm}This section formally defines the auction setting to be analyzed and the notations used throughout this paper. In particular, we compute and analyze the Bayesian Nash equilibrium for a \emph{multi-unit k-double auction (for $k=0.5$)} involving one buyer and one seller. The buyer is denoted by $B$, and the seller by $S$. The NE analysis is carried out for $m = 2$ units of identical and indivisible items/quantities.
        
        \hspace{0.5cm}Let us assume that the true types (true valuations) of buyer $B$ and seller $S$ are $\theta_B$ and $\theta_S$, respectively. Both $B$ and $S$ place two bids/asks in the auction by following \emph{scale-based bidding strategies} $b_B$ and $b_S$, respectively. The buyer $B's$ scale based strategy $b_B$ is defined as a strategy in which it places two bids $b_{B}^1 = \alpha_{B1}\theta_B$ and $b_{B}^2 = \alpha_{B2}\theta_B$; while seller $S$ places two asks $b_{S}^1 = \alpha_{S1}\theta_S$ and $b_{S}^2 = \alpha_{S2}\theta_S$ by following strategy $b_S$. Here, $\alpha_{B1}$ and $\alpha_{B2}$ are the scale factors by which buyer $B$ scales its true type (assuming $\alpha_{B1} \ge \alpha_{B2}$), while  $\alpha_{S1}$ and $\alpha_{S2}$ are the scale factors by which seller $S$ scales its true type (assuming $\alpha_{S1} \le \alpha_{S2}$). Additionally, we assume $\theta_B \sim U[l_B,h_B]$ and $\theta_S \sim U[l_S,h_S]$ and this is a common knowledge; which implies that the true types of the buyer and seller are sampled from a uniform distribution from the mentioned intervals. The expected utilities of B and S are denoted by $U_B$ and $U_S$, respectively. 
        
    %%%%%%%%%%%%%%%%%%%%%%%%%%%%%%%%%%%%%%%%%%%%%%%%%%%%%%%%%%%%%%%%%%%%%%%%

    \subsection{Bayesian Nash Equilibrium}
        \hspace{0.5cm}In this work, we are characterising an equilibrium in the \emph{game of incomplete information} as a player does not know the actual type of the opponent. Particularly, we are interested in Bayesian Nash equilibrium analysis.
        \begin{definition}[Bayesian Nash equilibrium]
            A Bayesian Nash equilibrium in a Bayesian game $\Gamma$ is defined as, 
            a profile of strategies $(s_1^*, s_2^*, ..., s_n^*)$ is a Bayesian Nash Equilibrium, if $\forall i \in N$; $\forall s_i : \phi_i \rightarrow S_i$; $\forall \theta_i \in \phi_i$, $$u_i((s_i^*, s_{-i}^*) \mid \theta_i) \ge u_i((s_i, s_{-i}^*) \mid \theta_i)$$
            That is, $\forall i \in N$; $\forall a_i \in S_i$; $\forall \theta_i \in \phi_i$,
            $$ \mathbb{E_{\theta_{-i}}} [u_i(\theta_i, \theta_{-i}, s_i^*(\theta_i), s_{-i}^*(\theta_{-i}))] \ge \mathbb{E_{\theta_{-i}}} [u_i(\theta_i, \theta_{-i}, a_i, s_{-i}^*(\theta_{-i}))] $$     
        where, $N$ is a set of players, $\theta_i$ and $u_i(.)$ are true type and utility of player $i$, respectively.
 \end{definition}
    %%%%%%%%%%%%%%%%%%%%%%%%%%%%%%%%%%%%%%%%%%%%%%%%%%%%%%%%%%%%%%%%%%%%%%%%

    \subsection{Markov Decision Process (MDP)}
        \hspace{0.5cm}We introduce a DDPG based bidding strategy for double auctions, which is an RL algorithm and follows the Markov Decision Process (MDP) framework,
        \begin{definition}[Markov Decision Process (MDP)]
            A Markov Decision Process (MDP) is a Reinforcement Learning framework; which is a tuple represented by $M = (S, A, P, r, \gamma)$, where $S$ denotes the set of states, $A$ denotes the set of actions, $P$ is the transition probability function, where $P(s'\mid s,a)=P(s_{t+1}=s' \mid s_t=s, a_t=a)$ is the probability that taking action $a$ in state $s$ at time $t$ will lead to state $s'$ at time t+1, $r$ denotes the reward function, where $r(s,a)$ is the reward obtained by taking action $a$ in state $s$, and $\gamma \in [0,1]$ denotes the discount factor.
        \end{definition}
        
    %%%%%%%%%%%%%%%%%%%%%%%%%%%%%%%%%%%%%%%%%%%%%%%%%%%%%%%%%%%%%%%%%%%%%%%%

    \subsection{PowerTAC} \label{ssec: powertac}
        \hspace{0.5cm}In our work, we focus on designing an optimal bidding strategy for PDA. PowerTAC~\cite{KETTER2013262} simulates the real-world smart-grid, which also includes wholesale energy market PDA. PowerTAC simulates a smart-grid for approximately 60 simulation days, wherein it replicates all major entities of smart-grids, i.e., various types of customers, power generation companies (GenCo), Transmission System Operator (TSO), and energy brokers. In any game in PowerTAC, multiple energy brokers compete against each other across three markets, namely wholesale, retail, and balancing market. A broker has to buy or sell energy in the wholesale market to keep supply and demand in balance. Here, we only focus on the wholesale and balancing markets to benchmark our DDPGBBS against state-of-the-art and baseline PowerTAC bidding strategies.
        
        \hspace{0.5cm}PowerTAC wholesale market PDA employs a day-ahead auction, wherein a broker can purchase energy 24 hours ahead of the delivery time. A total of 24 auctions take place periodically at one-hour intervals for any delivery timeslot. PowerTAC PDA exercises ACPR with uniform pricing as the payment rule. In wholesale market PDA, a broker has to compete with other brokers and the default players in the simulator. During a game, a broker does not know about other brokers' types (true valuations). For each hour of the game, a broker has to predict the demand and make bids in the wholesale PDA to fulfill the predicted demand. Each broker can participate in all 24 auctions and can submit any number of bids. The PowerTAC auction clearing mechanism clears the market after each periodic auction and communicates the relevant market-clearing information to each broker. Each broker is informed about the market-clearing price, its own cleared quantity, and net cleared quantity. The orderbook contains the anonymized list of uncleared bids and uncleared asks in the auction. Failing to purchase the required quantity from the wholesale market makes the broker purchase the remaining quantity from the balancing market by paying a balancing price that is generally higher than wholesale prices, which is meant to penalize the broker for creating an imbalance. We refer the reader to PowerTAC 2020 specifications to know more about the simulator~\cite{KETTER2013262}.
        
    \subsection*{Benchmark Strategies}
        The following are the PDA wholesale strategies used to benchmark DDPGBBS performance.
        
        \begin{itemize}
            \item Zero Intelligence (ZI): The ZI strategy follows a randomized approach to bid in a PDA by ignoring the state of the market; it samples a price from a uniform distribution between the minimum bid price and maximum bid price. Following the ZI strategy, the broker places one bid per auction and the remaining quantity as bid-quantity for all 24 auction instances. 
            \item Zero Intelligence Plus (ZIP): The ZIP agent~\cite{zip} maintains a scalar variable $m$ denoting the profit it aims to achieve, which gets combined with a unit limit price to compute a bid price $p$. Small increments adjust the price for each trade with the help of a $\delta$ by comparing the submitted bid price and the clearing price. We decide the bid price $\mu$ randomly at the start of the game. The profit margin $m$ is set to $-1\%$ of $\mu$, resulting in the initial bid price to be $p = \mu\  * \ 0.99$.
            \item VidyutVanika (VV): VidyutVanika~\cite{susobhan20,susobhan2} agent uses an MDP-based bidding strategy, where the MDP is solved with the help of dynamic programming and limit-price predictor. The MDP suggests a suitable limit-price for each of 24 auction instances. Additionally, it uses a heuristic strategy to spread the bid-quantity across 24 instances based on the predicted limit-prices.  
            \item SPOT: The SPOT~\cite{ijcai2018-23} agent uses a Monte Carlo Tree Search (MCTS) based strategy with coupled heuristics on top of the limit price derived from a REPTree based limit price predictor to determine the optimal bid price. This strategy places more than one bid in each auction instance by equally distributing the remaining required quantity among all the bids.
        \end{itemize}
%%%%%%%%%%%%%%%%%%%%%%%%%%%%%%%%%%%%%%%%%%%%%%%%%%%%%%%%%%%%%%%%%%%%%%%%

\section{Theoretical Analysis}\label{analysis}

    \hspace{0.5cm}Here, we show the theoretical analysis of double auctions. Consider a two-unit double auction involving one buyer and one seller, employing ACPR with uniform pricing as the payment rule. Both buyer and seller follow a scale-based bidding strategy $b_B$ and $b_S$, respectively. The buyer places two bids $b_B^1 = \alpha_{B1}\theta_B$ and $b_B^2 = \alpha_{B2}\theta_B$, while the seller places two asks $b_S^1 = \alpha_{S1}\theta_S$ and $b_S^2 = \alpha_{S2}\theta_S$. In the next section, we formalize the expected utilities of the buyer and the seller for a general two-unit double auction, where both the buyer and seller follow scale-based bidding strategies, and their types are drawn from $U[l_B,h_B]$ and $U[l_S,h_S]$, respectively.  

    %%%%%%%%%%%%%%%%%%%%%%%%%%%%%%%%%%%%%%%%%%%%%%%%%%%%%%%%%%%%%%%%%%%%%%%%
    
    \subsection{Nash Equilibrium Analysis for Multi-unit Double Auctions}
        
        \hspace{0.5cm}We start with an analysis of the buyer's expected utility. There are three possible clearing scenarios; We can write separate equations for each of the three scenarios and later combine them. $u_B^1$ is the expected utility when both units are cleared, $u_B^2$ is the expected utility when only one unit is cleared, and $u_B^3$ is the expected utility when no clearing happens. As the market does not clear in the third scenario, $u_B^3 = 0$. We can write them as follows:

        \begin{align}
            u_B^1 & = 2 \int_{l_S}^{\frac{\alpha_{B2}}{\alpha_{S2}}\theta_B} \left( \theta_B -  \left( \frac{\alpha_{B2}\theta_B + \alpha_{S2}\theta_S}{2} \right) \right) . \left( \frac{1}{h_S - l_S}\right) \ d\theta_S \numberthis \label{eq:ub1}
        \end{align}
        \begin{align}
            u_B^2 & = \int_{\frac{\alpha_{B2}}{\alpha_{S2}}\theta_B}^{\frac{\alpha_{B1}}{\alpha_{S1}}\theta_B} \left( \theta_B -  \left( \frac{\alpha_{B1}\theta_B + \alpha_{S1}\theta_S}{2} \right) \right) . \left( \frac{1}{h_S - l_S}\right) \ d\theta_S \numberthis \label{eq:ub2}
        \end{align}

        Assuming that the buyer decides to fix its $\alpha_{B1}$ and $\alpha_{B2}$ before even seeing its own type,

        \begin{align}
            U_B & = \int_{l_B}^{h_B} \left( u_B^1 + u_B^2 \right) . \left( \frac{1}{h_B - l_B} \right) \ d\theta_B \numberthis \label{eq:Ub}
        \end{align}
        Equations~\ref{eq:ub1} and \ref{eq:ub2} calculate the interim utility of the buyer for the scenarios involving two units of clearance and one unit clearance, respectively. The integral limits in Equations~\ref{eq:ub1} and \ref{eq:ub2} denote the regions where two units of clearance and one unit clearance happen, respectively. As we assume that the true types are sampled from the uniform distribution, the normalizing factors denote the probabilities of sampling from such a distribution. For the last case, when no clearing happens, the interim utility of the buyer is zero. Equation~\ref{eq:Ub} calculates the expected utility of the buyer based on the above three scenarios. In order to solve Equation~\ref{eq:Ub}, we first need to solve Equations~\ref{eq:ub1} and \ref{eq:ub2},

        \begin{align*}
            u_B^1 & = 2 \int_{l_S}^{\frac{\alpha_{B2}}{\alpha_{S2}}\theta_B} \left( \theta_B -  \left( \frac{\alpha_{B2}\theta_B + \alpha_{S2}\theta_S}{2} \right) \right) . \left( \frac{1}{h_S - l_S}\right) \ d\theta_S \\
            & = 2 \int_{l_S}^{\frac{\alpha_{B2}}{\alpha_{S2}}\theta_B} \left( \left( 1 -  \frac{\alpha_{B2}}{2} \right) \theta_B - \frac{\alpha_{S2}\theta_S}{2} \right) . \left( \frac{1}{h_S - l_S}\right) \ d\theta_S \\
            u_B^1 & = 2 \left( \left( 1 -  \frac{\alpha_{B2}}{2} \right) \left( {\frac{\alpha_{B2}}{\alpha_{S2}}\theta_B} - l_S\right) \theta_B - \frac{\alpha_{S2}}{4} \left( {\frac{\alpha_{B2}^2}{\alpha_{S2}^2}\theta_B^2} - l_S^2\right) \right) . \left( \frac{1}{h_S - l_S}\right) \numberthis \label{eq:ub1_ex} \\
            u_B^2 & = \int_{\frac{\alpha_{B2}}{\alpha_{S2}}\theta_B}^{\frac{\alpha_{B1}}{\alpha_{S1}}\theta_B} \left( \theta_B -  \left( \frac{\alpha_{B1}\theta_B + \alpha_{S1}\theta_S}{2} \right) \right) . \left( \frac{1}{h_S - l_S}\right) \ d\theta_S \\
            & = \int_{\frac{\alpha_{B2}}{\alpha_{S2}}\theta_B}^{\frac{\alpha_{B1}}{\alpha_{S1}}\theta_B} \left( \left( 1 -  \frac{\alpha_{B1}}{2} \right) \theta_B - \frac{\alpha_{S1}\theta_S}{2} \right) . \left( \frac{1}{h_S - l_S}\right) \ d\theta_S \\
            & = \left( \left( 1 -  \frac{\alpha_{B1}}{2} \right) \left( {\frac{\alpha_{B1}}{\alpha_{S1}}} -  \frac{\alpha_{B2}}{\alpha_{S2}} \right) \theta_B^2 - \frac{\alpha_{S1}}{4} \left( {\left( \frac{\alpha_{B1}}{\alpha_{S1}} \right)^2} -  \left( \frac{\alpha_{B2}}{\alpha_{S2}} \right)^2 \right) \theta_B^2 \right) . \left( \frac{1}{h_S - l_S}\right) \\
            u_B^2 & =  \theta_B^2 \left( {\frac{\alpha_{B1}}{\alpha_{S1}}} -  \frac{\alpha_{B2}}{\alpha_{S2}} \right) \left( 1 -  \frac{\alpha_{B1}}{2} - \frac{\alpha_{S1}}{4} \left( { \frac{\alpha_{B1}}{\alpha_{S1}}} +  \frac{\alpha_{B2}}{\alpha_{S2}} \right) \right) . \left( \frac{1}{h_S - l_S}\right) \numberthis \label{eq:ub2_ex}
        \end{align*}

        Solving Equation~\ref{eq:Ub} using the results in Equation~\ref{eq:ub1_ex} and \ref{eq:ub2_ex},
        \begin{align*}
            U_B & = \int_{l_B}^{h_B} \left( u_B^1 + u_B^2 \right) . \left( \frac{1}{h_B - l_B} \right) \ d\theta_B \\ 
            U_B & = \int_{l_B}^{h_B} \Bigg[ \Bigg. 2 \left( \left( 1 -  \frac{\alpha_{B2}}{2} \right) \left( {\frac{\alpha_{B2}}{\alpha_{S2}}\theta_B} - l_S\right) \theta_B - \frac{\alpha_{S2}}{4} \left( {\frac{\alpha_{B2}^2}{\alpha_{S2}^2}\theta_B^2} - l_S^2\right) \right) \\
            & + \left( {\frac{\alpha_{B1}}{\alpha_{S1}}} -  \frac{\alpha_{B2}}{\alpha_{S2}} \right) \theta_B^2 \left( 1 -  \frac{\alpha_{B1}}{2} - \frac{\alpha_{S1}}{4} \left( { \frac{\alpha_{B1}}{\alpha_{S1}}} +  \frac{\alpha_{B2}}{\alpha_{S2}} \right) \right) \Bigg] \Bigg. . \left( \frac{1}{h_S - l_S} \right) \left( \frac{1}{h_B - l_B} \right) \ d\theta_B \\
            U_B & = \Bigg[ \Bigg. 2 \left( \left( 1 -  \frac{\alpha_{B2}}{2} \right) \left( {\frac{\alpha_{B2}}{\alpha_{S2}} \left( \frac{h_B^3 - l_B^3}{3} \right)} - \frac{l_S(h_B^2 - l_B^2)}{2}\right) - \frac{\alpha_{S2}}{4} \left( {\frac{\alpha_{B2}^2}{\alpha_{S2}^2} \left( \frac{h_B^3 - l_B^3}{3} \right)} - l_S^2 (h_B - l_B)\right) \right) \\
            & + \left( {\frac{\alpha_{B1}}{\alpha_{S1}}} -  \frac{\alpha_{B2}}{\alpha_{S2}} \right)\left( 1 -  \frac{\alpha_{B1}}{2} - \frac{\alpha_{S1}}{4} \left( { \frac{\alpha_{B1}}{\alpha_{S1}}} +  \frac{\alpha_{B2}}{\alpha_{S2}} \right) \right) \left( \frac{h_B^3 - l_B^3}{3} \right) \Bigg] \Bigg. . \left( \frac{1}{h_S - l_S} \right) \left( \frac{1}{h_B - l_B} \right)\\
            U_B & = \left( \frac{h_B^3 - l_B^3}{3(h_B - l_B)(h_S - l_S)} \right)  \Bigg[ \Bigg. 2 \left( \left( 1 -  \frac{\alpha_{B2}}{2} \right) {\frac{\alpha_{B2}}{\alpha_{S2}}} - \frac{\alpha_{B2}^2}{4\alpha_{S2}} \right) + \left( {\frac{\alpha_{B1}}{\alpha_{S1}}} -  \frac{\alpha_{B2}}{\alpha_{S2}} \right)\left( 1 -  \frac{\alpha_{B1}}{2} - \frac{\alpha_{S1}}{4} \left( { \frac{\alpha_{B1}}{\alpha_{S1}}} +  \frac{\alpha_{B2}}{\alpha_{S2}} \right) \right) \Bigg] \Bigg. \\
            & - l_S\frac{(h_B^2 - l_B^2)}{(h_B - l_B)(h_S - l_S)} \left(1 - \frac{\alpha_{B2}}{2} \right) + \frac{\alpha_{S2}l_S^2}{2}\frac{(h_B - l_B)}{(h_B - l_B)(h_S - l_S)}\\
            U_B & = \frac{(h_B^2 + h_Bl_B + l_B^2)}{3(h_S - l_S)} \Bigg( \Bigg. \left( \frac{\alpha_{B1}}{\alpha_{S1}} - \frac{\alpha_{B2}}{\alpha_{S2}} \right) \left( 1 - \frac{\alpha_{B1}}{2} - \frac{\alpha_{S1}}{4} \left( \frac{\alpha_{B1}}{\alpha_{S1}} + \frac{\alpha_{B2}}{\alpha_{S2}} \right) \right) 
            + 2\frac{\alpha_{B2}}{\alpha_{S2}} \left( 1 - \frac{\alpha_{B2}}{2} \right) - \frac{\alpha_{B2}^2}{2\alpha_{S2}} \Bigg) \Bigg. \\
            & - l_S\frac{(h_B + l_B)}{(h_S - l_S)} \left( 1 - \frac{\alpha_{B2}}{2} \right) + \frac{1}{2(h_S - l_S)}l_S^2\alpha_{S2} \numberthis \label{eq:Ub_extended}
        \end{align*} 
        \\
        We now calculate the seller's expected utility following the similar logic as above,

        \begin{align}
        u_S^1 & = 2 \int_{\frac{\alpha_{S2}\theta_S}{\alpha_{B2}}}^{h_B} \left( \left( \frac{\alpha_{B2}\theta_B + \alpha_{S2}\theta_S}{2} \right) - \theta_S \right) . \left( \frac{1}{h_B - l_B} \right) \ d\theta_B \numberthis \label{eq:us1}
        \end{align}
        \begin{align}
        u_S^2 & = \int_{\frac{\alpha_{S1}\theta_S}{\alpha_{B1}}}^{\frac{\alpha_{S2}\theta_S}{\alpha_{B2}}} \left( \left( \frac{\alpha_{B1}\theta_B + \alpha_{S1}\theta_S}{2} \right) - \theta_S \right) . \left( \frac{1}{h_B - l_B} \right) \ d\theta_B \numberthis \label{eq:us2}
        \end{align}

        Assuming that the seller decides to fix its $\alpha_{S1}$ and $\alpha_{S2}$ before even seeing its own type,

        \begin{align}
        U_S & = \int_{l_S}^{h_S} \left( u_S^1 + u_S^2 \right) . \left( \frac{1}{h_S - l_S} \right) \ d\theta_S \numberthis \label{eq:Us}
        \end{align}
        Solving the above equations would result in,

        \begin{align*}
            U_S & = \frac{(h_S^2 + h_Sl_S + l_S^2)}{3(h_B - l_B)} \Bigg( \Bigg. \left( \frac{\alpha_{S2}}{\alpha_{B2}} - \frac{\alpha_{S1}}{\alpha_{B1}} \right) \left( \frac{\alpha_{B1}}{4} \left( \frac{\alpha_{S2}}{\alpha_{B2}} + \frac{\alpha_{S1}}{\alpha_{B1}} \right) + \frac{\alpha_{S1}}{2} - 1 \right) 
            + 2\frac{\alpha_{S2}}{\alpha_{B2}} \left( 1 - \frac{\alpha_{S2}}{2} \right) - \frac{\alpha_{S2}^2}{2\alpha_{B2}} \Bigg) \Bigg. \\
            & - h_B\frac{(h_S + l_S)}{(h_B - l_B)}\left( 1 - \frac{\alpha_{S2}}{2} \right) + \frac{1}{2(h_B - l_B)}h_B^2\alpha_{B2} \numberthis \label{eq:Us_extended}
        \end{align*}
        \\
        Now, We consider the following four possible scenarios based on the scale factors of the buyer and the seller: \\
        \textsc{Case-1}: Same Scale-Factors for Buyer, Same Scale Factors for Seller \\
        \textsc{Case-2}: Different Scale Factors for Buyer, Same Scale Factors for Seller \\
        \textsc{Case-3}: Same Scale Factors for Buyer, Different Scale Factors for Seller \\
        \textsc{Case-4}: Different Scale Factors for Buyer, Different Scale Factors for Seller \\
        
        %%%%%%%%%%%%%%%%%%%%%%%%%%%%%%%%%%%%%%%%%%%%%%%%%%%%%%%%%%%%%%%%%%%%%%%%
        
        \subsubsection{\textsc{Case-1: }\textbf{Same Scale Factors for Buyer, Same Scale Factors for Seller}}
        
           Here $\alpha_{B1} = \alpha_{B2} = \alpha_B$ and $\alpha_{S1} = \alpha_{S2} = \alpha_S$. Rewriting Equations~\ref{eq:Ub_extended} and \ref{eq:Us_extended} for this case,

            \begin{align*}
                U_B & = \frac{(h_B^2 + h_Bl_B + l_B^2)}{6\alpha_S(h_S - l_S)} (\alpha_B (4 - 3 \alpha_B)) - l_S\frac{(h_B + l_B)}{(h_S - l_S)} \left( 1 - \frac{\alpha_{B}}{2} \right) + \frac{1}{2(h_S - l_S)}l_S^2\alpha_{S} \numberthis \label{eq:Ub_case1}
            \end{align*} 
            \begin{align*}
                U_S & = \frac{(h_S^2 + h_Sl_S + l_S^2)}{6\alpha_B(h_B - l_B)} (\alpha_S (4 - 3 \alpha_S)) - h_B\frac{(h_S + l_S)}{(h_B - l_B)}\left( 1 - \frac{\alpha_{S}}{2} \right) + \frac{1}{2(h_B - l_B)}h_B^2\alpha_{B} \numberthis \label{eq:Us_case1}
            \end{align*}

           Now, in order to find the $\alpha_B$ that maximizes the buyer's expected utility $U_B$, we take $\frac{\partial{U_B}}{\partial{\alpha_B}} = 0$, which implies

           \begin{align*}
               & \frac{(h_B^2 + h_Bl_B + l_B^2)}{6\alpha_S(h_S - l_S)} (4 - 6\alpha_B) + \frac{l_S}{2} \left( \frac{h_B + l_B}{h_S - l_S} \right) = 0 \Rightarrow \\
               & \frac{(h_B^2 + h_Bl_B + l_B^2)}{(h_B + l_B)} (4 - 6\alpha_B) + 3l_S\alpha_S = 0 \Rightarrow \\
               & 4 - 6\alpha_B = -3l_S\alpha_S \left( \frac{h_B + l_B}{h_B^2 + h_Bl_B + l_B^2} \right) \Rightarrow \\
               & \alpha_B = \frac{2}{3} + \frac{l_S\alpha_S}{2} \left( \frac{h_B + l_B}{h_B^2 + h_Bl_B + l_B^2} \right) \numberthis \label{eq:case1_alphab}
           \end{align*}

           To find the $\alpha_S$ that maximizes the seller's expected utility $U_S$, we take $\frac{\partial{U_S}}{\partial{\alpha_S}} = 0$, which implies

           \begin{align*}
               & \frac{(h_S^2 + h_Sl_S + l_S^2)}{6\alpha_B(h_B - l_B)} (4 - 6\alpha_S) + \frac{h_B}{2}\frac{(h_S + l_S)}{(h_B - l_B)} = 0 \Rightarrow \\
               & \frac{(h_S^2 + h_Sl_S + l_S^2)}{(h_S + l_S)} (4 - 6\alpha_S) + 3h_B\alpha_B = 0 \Rightarrow \\
               & 4 - 6\alpha_S = -3h_B\alpha_B \left( \frac{h_S + l_S}{h_S^2 + h_Sl_S + l_S^2} \right) \Rightarrow \\
               & \alpha_S = \frac{2}{3} + \frac{h_B\alpha_B}{2} \left( \frac{h_S + l_S}{h_S^2 + h_Sl_S + l_S^2} \right) \numberthis \label{eq:case1_alphas}
           \end{align*}

           Above $\alpha_B$ and $\alpha_S$ calculate scale factors for a generic case when $\theta_B \sim U[l_B,h_B]$ and $\theta_S \sim U[l_S,h_S]$. 
           
           \begin{remark}
               For a specific case of $\theta_B \sim U[0,1]$ and $\theta_S \sim U[0,1]$, replacing $l_B = l_S = 0$ and $h_S = h_B = 1$ in the above equations would result in $\alpha_B = \frac{2}{3}$ and $\alpha_S = 1$. 
           \end{remark}

        %%%%%%%%%%%%%%%%%%%%%%%%%%%%%%%%%%%%%%%%%%%%%%%%%%%%%%%%%%%%%%%%%%%%%%%%
        
        \subsubsection{\textsc{Case-2: }\textbf{Different Scale Factors for Buyer, Same Scale Factors for Seller}}
        
            Here $\alpha_{S1} = \alpha_{S2} = \alpha_S$. Rewriting Equations~\ref{eq:Ub_extended} and \ref{eq:Us_extended} for this case,

            \begin{align*}
                U_B & = \frac{(h_B^2 + h_Bl_B + l_B^2)}{3\alpha_S(h_S - l_S)} \Bigg( \Bigg. ( \alpha_{B1} - \alpha_{B2}) \left( 1 - \frac{3\alpha_{B1}}{4} - \frac{\alpha_{B2}}{4} \right) - \frac{3\alpha_{B2}^2}{2} 
                + 2{\alpha_{B2}} \Bigg) \Bigg. - l_S\frac{(h_B + l_B)}{(h_S - l_S)} \left( 1 - \frac{\alpha_{B2}}{2} \right) \\
                & + \frac{1}{2(h_S - l_S)}l_S^2\alpha_{S} \numberthis \label{eq:Ub_case2} \\
                U_S & = \frac{(h_S^2 + h_Sl_S + l_S^2)}{3(h_B - l_B)} \Bigg( \Bigg. \alpha_{S} \left( \frac{1}{\alpha_{B2}} - \frac{1}{\alpha_{B1}} \right) \left( \frac{\alpha_{S}}{4} \left( \frac{\alpha_{B1}}{\alpha_{B2}} + 1 \right) + \frac{\alpha_{S}}{2} - 1 \right)
                +\frac{\alpha_{S}}{\alpha_{B2}} \left( 2 - \frac{3\alpha_{S}}{2} \right) \Bigg) \Bigg. \\
                & - h_B\frac{(h_S + l_S)}{(h_B - l_B)}\left( 1 - \frac{\alpha_{S2}}{2} \right) + \frac{1}{2(h_B - l_B)}h_B^2\alpha_{B2} \numberthis \label{eq:Us_case2}
            \end{align*}

            Now, to find the $\alpha_{B1}$ that maximizes the buyer's expected utility $U_B$, we take $\frac{\partial{U_B}}{\partial{\alpha_{B1}}} = 0$, which implies,

           \begin{align*}
                &(\alpha_{B1} - \alpha_{B2})(\frac{-3}{4}) + \left( 1 - \frac{3\alpha_{B1}}{4} - \frac{\alpha_{B2}}{4} \right) = 0 \Rightarrow \\
                & (-3)(\alpha_{B1} - \alpha_{B2}) + (4 - 3\alpha_{B1} - \alpha_{B2}) = 0 \Rightarrow \\
                &3\alpha_{B1} - \alpha_{B2} - 2 = 0 \numberthis \label{eq:case2_gen1}
            \end{align*}

           To find the $\alpha_{B2}$ that maximizes $U_B$, we take $\frac{\partial{U_B}}{\partial{\alpha_{B2}}} = 0 $, which implies,

           \begin{align*}
               &\frac{(h_B^2 + h_Bl_B + l_B^2)}{3\alpha_S(h_S - l_S)} \Bigg( \Bigg. ( \alpha_{B1} - \alpha_{B2}) \left( \frac{-1}{4} \right) - \left( 1 - \frac{3\alpha_{B1}}{4} - \frac{\alpha_{B2}}{4} \right) - 3\alpha_{B2} + 2 \Bigg) \Bigg. - l_S\frac{(h_B + l_B)}{(h_S - l_S)} \left( \frac{-1}{2} \right) = 0 \Rightarrow \\
               &((-1)(\alpha_{B1} - \alpha_{B2}) - (4 - 3\alpha_{B1} - \alpha_{B2}) - 12\alpha_{B2} + 8) \left( \frac{h_B^2 + h_Bl_B + l_B^2}{h_B + l_B} \right) + 6l_S\alpha_S = 0 \Rightarrow \\
               &(\alpha_{B1} - 5\alpha_{B2} + 2) \left( \frac{h_B^2 + h_Bl_B + l_B^2}{h_B + l_B} \right) + 3l_S\alpha_S = 0 \numberthis \label{eq:case2_gen2}
           \end{align*}

           To find the $\alpha_{S}$ that maximizes $U_S$, we take $\frac{\partial{U_S}}{\partial{\alpha_S}} = 0$, which implies

           \begin{align*}
               & \frac{(h_S^2 + h_Sl_S + l_S^2)}{3(h_B - l_B)} \Bigg( \Bigg. \alpha_{S} \left( \frac{1}{\alpha_{B2}} - \frac{1}{\alpha_{B1}} \right) \left( \frac{1}{4} \left( \frac{\alpha_{B1}}{\alpha_{B2}} + 1 \right) + \frac{1}{2} \right) + \left( \frac{1}{\alpha_{B2}} - \frac{1}{\alpha_{B1}} \right) \left( \frac{\alpha_{S}}{4} \left( \frac{\alpha_{B1}}{\alpha_{B2}} + 1 \right) + \frac{\alpha_{S}}{2} - 1 \right) \\
               & + \frac{1}{\alpha_{B2}} \left( 2 - 3\alpha_{S} \right) \Bigg) \Bigg. - h_B\frac{(h_S + l_S)}{(h_B - l_B)}\left(\frac{-1}{2} \right) = 0 \Rightarrow \\
               & \Bigg( \Bigg.\Bigg( \Bigg. \frac{1}{\alpha_{B2}} - \frac{1}{\alpha_{B1}} \Bigg) \Bigg. \Bigg( \Bigg. \frac{\alpha_S}{2} \Bigg( \Bigg. \frac{\alpha_{B1}}{\alpha_{B2}} + 3 \Bigg) \Bigg. - 1 \Bigg) \Bigg. + \frac{1}{\alpha_{B2}} (2 - 3\alpha_S) \Bigg) \Bigg. \left(\frac{h_S^2 + h_Sl_S + l_S^2}{h_S + l_S} \right) + \frac{3}{2}h_B = 0 \Rightarrow \\
               & \Bigg( \Bigg.\Bigg( \Bigg. \frac{1}{\alpha_{B2}} - \frac{1}{\alpha_{B1}} \Bigg) \Bigg. (\alpha_S (\alpha_{B1} + 3\alpha_{B2}) - 2\alpha_{B2}) - 6\alpha_S + 4 \Bigg) \Bigg. \left(\frac{h_S^2 + h_Sl_S + l_S^2}{h_S + l_S} \right) + 3h_B\alpha_{B2} = 0 \numberthis \label{eq:case2_gen3}
           \end{align*}

           \begin{remark}
               For a specific case of $\theta_B \sim U[0,1]$ and $\theta_S \sim U[0,1]$, replacing $l_B = l_S = 0$ and $h_S = h_B = 1$ in the above equations, we get the following set of equations,
           
               \begin{align*}
                   & 3\alpha_{B1} - \alpha_{B2} - 2 = 0 \numberthis \label{eq:case2_1} \\
                   & \alpha_{B1} - 5\alpha_{B2} + 2 = 0 \numberthis \label{eq:case2_2} \\
                   & \Bigg( \Bigg. \frac{1}{\alpha_{B2}} - \frac{1}{\alpha_{B1}} \Bigg) \Bigg. (\alpha_S (\alpha_{B1} + 3\alpha_{B2}) - 2\alpha_{B2}) - 6\alpha_S + 4 + 3\alpha_{B2} = 0 \numberthis \label{eq:case2_3}
               \end{align*}
               
               By solving these equations, we get $\alpha_{B1} = \frac{6}{7}$, $\alpha_{B2} = \frac{4}{7}$ and $\alpha_S = \frac{268}{315}$. 
           \end{remark}
        
        %%%%%%%%%%%%%%%%%%%%%%%%%%%%%%%%%%%%%%%%%%%%%%%%%%%%%%%%%%%%%%%%%%%%%%%%
        
        \subsubsection{\textsc{Case-3: }\textbf{Same Scale Factors for Buyer, Different Scale Factors for Seller}} 
        
        Here $\alpha_{B1} = \alpha_{B2} = \alpha_B$. Rewriting Equations~\ref{eq:Ub_extended} and \ref{eq:Us_extended} for this case,

            \begin{align*}
                &U_B = \frac{(h_B^2 + h_Bl_B + l_B^2)}{3(h_S - l_S)} \Bigg( \Bigg. \alpha_{B} \left( \frac{1}{\alpha_{S1}} - \frac{1}{\alpha_{S2}} \right) \left( 1 - \frac{\alpha_{B}}{2} - \frac{\alpha_{B}}{4} \left( 1 + \frac{\alpha_{S1}}{\alpha_{S2}} \right) \right) 
                + \frac{\alpha_{B}}{\alpha_{S2}} \left( 2 - \frac{3\alpha_{B}}{2} \right) \Bigg) \Bigg. \\
                & - l_S\frac{(h_B + l_B)}{(h_S - l_S)} \left( 1 - \frac{\alpha_{B}}{2} \right) + \frac{1}{2(h_S - l_S)}l_S^2\alpha_{S2} \numberthis \label{eq:Ub_case3} \\
                &U_S = \frac{(h_S^2 + h_Sl_S + l_S^2)}{3\alpha_B(h_B - l_B)} \Bigg( \Bigg. (\alpha_{S2} - \alpha_{S1}) \left( \frac{3\alpha_{S1}}{4} + \frac{\alpha_{S2}}{4} -1 \right) - \frac{3\alpha_{S2}^2}{2} + 2\alpha_{S2} \Bigg) \Bigg. - h_B\frac{(h_S + l_S)}{(h_B - l_B)}\left( 1 - \frac{\alpha_{S2}}{2} \right) \\ 
                & + \frac{1}{2(h_B - l_B)}h_B^2\alpha_{B} \numberthis \label{eq:Us_case3}
            \end{align*}

            Now, to find the $\alpha_{B}$ that maximizes the buyer's expected utility $U_B$, we take $\frac{\partial{U_B}}{\partial{\alpha_{B}}} = 0$, which implies,

           \begin{align*}
                &\frac{(h_B^2 + h_Bl_B + l_B^2)}{3(h_S - l_S)} \Bigg( \Bigg. \left( \frac{1}{\alpha_{S1}} - \frac{1}{\alpha_{S2}} \right) \left( 1 - \alpha_{B} - \frac{\alpha_{B}}{2} \left( 1 + \frac{\alpha_{S1}}{\alpha_{S2}} \right) \right) 
                + \frac{1}{\alpha_{S2}} \left( 2 - 3\alpha_{B} \right) \Bigg) \Bigg. + \frac{l_S}{2}\frac{(h_B + l_B)}{(h_S - l_S)} = 0 \Rightarrow \\
                &\Bigg( \Bigg. \Bigg( \Bigg. \frac{1}{\alpha_{S1}} - \frac{1}{\alpha_{S2}} \Bigg) \Bigg. (2\alpha_{S2} (1 - \alpha_B) - \alpha_B (\alpha_{S2} + \alpha_{S1})) \Bigg. + 4 - 6\alpha_B \Bigg) \Bigg. \Bigg( \frac{1}{2\alpha_{S2}} \Bigg) \Bigg. \Bigg. \left(\frac{h_B^2 + h_Bl_B + l_B^2}{h_B + l_B} \right) + \frac{3}{2} l_S = 0 \Rightarrow \\
                &\Bigg( \Bigg. \Bigg( \Bigg. \frac{1}{\alpha_{S2}} - \frac{1}{\alpha_{S1}} \Bigg) \Bigg. (\alpha_B (\alpha_{S1} + 3\alpha_{S2}) - 2\alpha_{S2}) - 6\alpha_B + 4 \Bigg) \Bigg. \left(\frac{h_B^2 + h_Bl_B + l_B^2}{h_B + l_B} \right) + 3l_S\alpha_{S2} = 0 \numberthis \label{eq:case3_gen1} \\
            \end{align*}

           To find the $\alpha_{S1}$ that maximizes $U_S$, we take $\frac{\partial{U_S}}{\partial{\alpha_{S1}}} = 0 $, which implies,

           \begin{align*}
               &\frac{(h_S^2 + h_Sl_S + l_S^2)}{3\alpha_B(h_B - l_B)} \Bigg( \Bigg. (-1) \left( \frac{3\alpha_{S1}}{4} + \frac{\alpha_{S2}}{4} -1 \right) + (\alpha_{S2} - \alpha_{S1}) \left( \frac{3}{4} \right) \Bigg) \Bigg. = 0 \Rightarrow \\
               &(-1)(3\alpha_{S1} + \alpha_{S2} -4) + 3(\alpha_{S2} - \alpha_{S1}) = 0 \Rightarrow \\
               &3\alpha_{S1} - \alpha_{S2} - 2 = 0 \numberthis \label{eq:case3_gen2}
           \end{align*}

           To find the $\alpha_{S2}$ that maximizes $U_S$, we take $\frac{\partial{U_S}}{\partial{\alpha_S2}} = 0$, which implies

           \begin{align*}
               &\frac{(h_S^2 + h_Sl_S + l_S^2)}{3\alpha_B(h_B - l_B)} \Bigg( \Bigg. \left( \frac{3\alpha_{S1}}{4} + \frac{\alpha_{S2}}{4} -1 \right) + (\alpha_{S2} - \alpha_{S1}) \left( \frac{1}{4} \right) - 3\alpha_{S2} + 2 \Bigg) \Bigg. - h_B\frac{(h_S + l_S)}{(h_B - l_B)}\left( -\frac{1}{2} \right) = 0 \Rightarrow \\
               &((3\alpha_{S1} + \alpha_{S2} - 4) + (\alpha_{S2} - \alpha_{S1}) - 12\alpha_{S2} + 8)\left( \frac{h_S^2 + h_Sl_S + l_S^2}{h_S + l_S} \right) + 6h_B\alpha_B = 0\Rightarrow \\
               &(\alpha_{S1} - 5\alpha_{S2} + 2)\left( \frac{h_S^2 + h_Sl_S + l_S^2}{h_S + l_S} \right) + 3h_B\alpha_B = 0 \numberthis \label{eq:case3_gen3}
           \end{align*}

         \begin{remark}
            For a specific case of $\theta_B \sim U[0,1]$ and $\theta_S \sim U[0,1]$, replacing $l_B = l_S = 0$ and $h_S = h_B = 1$ in the above equations, we get the following set of equations,
       
           \begin{align*}
               & \Bigg( \Bigg. \frac{1}{\alpha_{S2}} - \frac{1}{\alpha_{S1}} \Bigg) \Bigg. (\alpha_B (\alpha_{S1} + 3\alpha_{S2}) - 2\alpha_{S2}) - 6\alpha_B + 4 = 0 \numberthis \label{eq:case3_1} \\
               & 3\alpha_{S1} - \alpha_{S2} - 2 = 0 \numberthis \label{eq:case3_2} \\
               & \alpha_{S1} - 5\alpha_{S2} + 2 + 3\alpha_B = 0 \numberthis \label{eq:case3_3}
           \end{align*}
           
           By solving these equations, we get $\alpha_{B} = \frac{2}{3}$, $\alpha_{S1} = 1$ and $\alpha_{S2} = 1$.
         \end{remark}
               
        %%%%%%%%%%%%%%%%%%%%%%%%%%%%%%%%%%%%%%%%%%%%%%%%%%%%%%%%%%%%%%%%%%%%%%%%
        
        \subsubsection{\textsc{Case-4: }\textbf{Different Scale Factors for Buyer, Different Scale Factors for Seller}}
        
            Recalling Equations~\ref{eq:Ub_extended} and \ref{eq:Us_extended} for this case,

            \begin{align*}
                U_B & = \frac{(h_B^2 + h_Bl_B + l_B^2)}{3(h_S - l_S)} \Bigg( \Bigg. \left( \frac{\alpha_{B1}}{\alpha_{S1}} - \frac{\alpha_{B2}}{\alpha_{S2}} \right) \left( 1 - \frac{\alpha_{B1}}{2} - \frac{\alpha_{S1}}{4} \left( \frac{\alpha_{B1}}{\alpha_{S1}} + \frac{\alpha_{B2}}{\alpha_{S2}} \right) \right) 
                + 2\frac{\alpha_{B2}}{\alpha_{S2}} \left( 1 - \frac{\alpha_{B2}}{2} \right) - \frac{\alpha_{B2}^2}{2\alpha_{S2}} \Bigg) \Bigg. \\
                & - l_S\frac{(h_B + l_B)}{(h_S - l_S)} \left( 1 - \frac{\alpha_{B2}}{2} \right) + \frac{1}{2(h_S - l_S)}l_S^2\alpha_{S2} 
            \end{align*} 
            \begin{align*}
                U_S & = \frac{(h_S^2 + h_Sl_S + l_S^2)}{3(h_B - l_B)} \Bigg( \Bigg. \left( \frac{\alpha_{S2}}{\alpha_{B2}} - \frac{\alpha_{S1}}{\alpha_{B1}} \right) \left( \frac{\alpha_{B1}}{4} \left( \frac{\alpha_{S2}}{\alpha_{B2}} + \frac{\alpha_{S1}}{\alpha_{B1}} \right) + \frac{\alpha_{S1}}{2} - 1 \right) 
                + 2\frac{\alpha_{S2}}{\alpha_{B2}} \left( 1 - \frac{\alpha_{S2}}{2} \right) - \frac{\alpha_{S2}^2}{2\alpha_{B2}} \Bigg) \Bigg. \\
                & - h_B\frac{(h_S + l_S)}{(h_B - l_B)}\left( 1 - \frac{\alpha_{S2}}{2} \right) + \frac{1}{2(h_B - l_B)}h_B^2\alpha_{B2}
            \end{align*}

            Now, to find the $\alpha_{B1}$ that maximizes the buyer's expected utility $U_B$, we take $\frac{\partial{U_B}}{\partial{\alpha_{B1}}} = 0$, which implies,

           \begin{align*}
                &\frac{(h_B^2 + h_Bl_B + l_B^2)}{3(h_S - l_S)} \Bigg( \Bigg. \left( \frac{\alpha_{B1}}{\alpha_{S1}} - \frac{\alpha_{B2}}{\alpha_{S2}} \right) \left( \frac{-3}{4} \right) + \left( \frac{1}{\alpha_{S1}} \right) \left( 1 - \frac{\alpha_{B1}}{2} - \frac{\alpha_{S1}}{4} \left( \frac{\alpha_{B1}}{\alpha_{S1}} + \frac{\alpha_{B2}}{\alpha_{S2}} \right) \right) \Bigg) \Bigg. = 0 \Rightarrow \\
                &\left( \frac{\alpha_{B1}}{\alpha_{S1}} - \frac{\alpha_{B2}}{\alpha_{S2}} \right) \left( \frac{-3}{4} \right) + \left( \frac{1}{\alpha_{S1}} - \frac{\alpha_{B1}}{2\alpha_{S1}} - \frac{1}{4} \left( \frac{\alpha_{B1}}{\alpha_{S1}} + \frac{\alpha_{B2}}{\alpha_{S2}} \right) \right) = 0 \Rightarrow \\
                &-3\alpha_{B1}\alpha_{S2} + 3\alpha_{B2}\alpha_{S1} + 4\alpha_{S2} - 2\alpha_{B1}\alpha_{S2} - \alpha_{B1}\alpha_{S2} - \alpha_{B2}\alpha_{S1} = 0 \Rightarrow \\
                &2\alpha_{B2}\alpha_{S1} + 4\alpha_{S2} - 6\alpha_{B1}\alpha_{S2} = 0 \Rightarrow \\
                &\alpha_{B2}\alpha_{S1} + \alpha_{S2} (2 - 3\alpha_{B1}) = 0 \numberthis \label{eq:case4_gen1}
            \end{align*}

           Then, to find the $\alpha_{B2}$ that maximizes the buyer's expected utility $U_B$, we take $\frac{\partial{U_B}}{\partial{\alpha_{B2}}} = 0$, which implies,

           \begin{align*}
                &\frac{(h_B^2 + h_Bl_B + l_B^2)}{3(h_S - l_S)} \Bigg( \Bigg. \left( \frac{\alpha_{B1}}{\alpha_{S1}} - \frac{\alpha_{B2}}{\alpha_{S2}} \right) \left( \frac{-\alpha_{S1}}{4\alpha_{S2}} \right) + \left( - \frac{1}{\alpha_{S2}} \right) \left( 1 - \frac{\alpha_{B1}}{2} - \frac{\alpha_{S1}}{4} \left( \frac{\alpha_{B1}}{\alpha_{S1}} + \frac{\alpha_{B2}}{\alpha_{S2}} \right) \right) \\
                &+ \frac{2}{\alpha_{S2}} - \frac{3\alpha_{B2}}{\alpha_{S2}} \Bigg) \Bigg. - l_S\frac{(h_B + l_B)}{(h_S - l_S)} \left( \frac{-1}{2}\right) = 0 \Rightarrow \\
                &\left( \frac{h_B^2 + h_Bl_B + l_B^2}{h_B + l_B} \right) \Bigg( \Bigg. \left( \frac{-\alpha_{B1}}{4} + \frac{\alpha_{B2}\alpha_{S1}}{4\alpha_{S2}} \right)- \left( 1 - \frac{3\alpha_{B1}}{4} - \frac{\alpha_{S1}\alpha_{B2}}{4\alpha_{S2}} \right) + 2 - 3\alpha_{B2} \Bigg) \Bigg. + \frac{3}{2}l_S\alpha_{S2} = 0 \Rightarrow \\
                &\left( \frac{h_B^2 + h_Bl_B + l_B^2}{h_B + l_B} \right) \Bigg( \Bigg. \frac{\alpha_{B1}}{2} + \frac{\alpha_{B2}\alpha_{S1}}{2\alpha_{S2}} + 1 - 3\alpha_{B2} \Bigg) \Bigg. + \frac{3}{2}l_S\alpha_{S2} = 0 \Rightarrow \\
                &\Bigg( \Bigg. 2 + \alpha_{B1} - 6\alpha_{B2} +  \frac{\alpha_{S1}\alpha_{B2}}{\alpha_{S2}} \Bigg) \Bigg. \left( \frac{h_B^2 + h_Bl_B + l_B^2}{h_B + l_B} \right) + 3l_S\alpha_{S2} = 0 \numberthis \label{eq:case4_gen2}
            \end{align*}

           Now, To find the $\alpha_{S1}$ that maximizes $U_S$, we take $\frac{\partial{U_S}}{\partial{\alpha_{S1}}} = 0 $, which implies,

           \begin{align*}
               &\frac{(h_S^2 + h_Sl_S + l_S^2)}{3(h_B - l_B)} \Bigg( \Bigg. \left( \frac{\alpha_{S2}}{\alpha_{B2}} - \frac{\alpha_{S1}}{\alpha_{B1}} \right) \left( \frac{3}{4} \right) + \left(- \frac{1}{\alpha_{B1}} \right) \left( \frac{\alpha_{B1}}{4} \left( \frac{\alpha_{S2}}{\alpha_{B2}} + \frac{\alpha_{S1}}{\alpha_{B1}} \right) + \frac{\alpha_{S1}}{2} - 1 \right) \Bigg) \Bigg. = 0 \Rightarrow \\
               &\left( \frac{\alpha_{S2}}{\alpha_{B2}} - \frac{\alpha_{S1}}{\alpha_{B1}} \right) \left( \frac{3}{4} \right) - \left( \frac{1}{4} \left( \frac{\alpha_{S2}}{\alpha_{B2}} + \frac{\alpha_{S1}}{\alpha_{B1}} \right) + \frac{\alpha_{S1}}{2\alpha_{B1}} - \frac{1}{\alpha_{B1}} \right) = 0 \Rightarrow \\
               &3\alpha_{B1}\alpha_{S2} - 3\alpha_{B2}\alpha_{S1} - \alpha_{B1}\alpha_{S2} - \alpha_{B2}\alpha_{S1} - 2\alpha_{B2}\alpha_{S1} + 4\alpha_{B2} = 0 \Rightarrow \\
               &2\alpha_{B1}\alpha_{S2} + 4\alpha_{B2} - 6\alpha_{B2}\alpha_{S1} = 0 \Rightarrow \\
               &\alpha_{B1}\alpha_{S2} + \alpha_{B2}(2 - 3\alpha_{S1}) = 0 \numberthis \label{eq:case4_gen3}
           \end{align*}

           Finally, To find the $\alpha_{S2}$ that maximizes $U_S$, we take $\frac{\partial{U_S}}{\partial{\alpha_{S2}}} = 0$, which implies

           \begin{align*}
               &\frac{(h_S^2 + h_Sl_S + l_S^2)}{3(h_B - l_B)} \Bigg( \Bigg. \left( \frac{\alpha_{S2}}{\alpha_{B2}} - \frac{\alpha_{S1}}{\alpha_{B1}} \right) \left( \frac{\alpha_{B1}}{4\alpha_{B2}} \right) + \left( \frac{1}{\alpha_{B2}} \right) \left( \frac{\alpha_{B1}}{4} \left( \frac{\alpha_{S2}}{\alpha_{B2}} + \frac{\alpha_{S1}}{\alpha_{B1}} \right) + \frac{\alpha_{S1}}{2} - 1 \right) \\
               &+ \frac{2}{\alpha_{B2}} - \frac{3\alpha_{S2}}{\alpha_{B2}} \Bigg) \Bigg. - h_B\frac{(h_S + l_S)}{(h_B - l_B)}\left(\frac{-1}{2} \right) = 0 \Rightarrow \\
               &\left(\frac{h_S^2 + h_Sl_S + l_S^2}{h_S + l_S)} \right) \Bigg( \Bigg. \left( \frac{\alpha_{B1}\alpha_{S2}}{4\alpha_{B2}} - \frac{\alpha_{S1}}{4} \right) + \left(  \frac{\alpha_{B1}\alpha_{S2}}{4\alpha_{B2}} + \frac{3\alpha_{S1}}{4}- 1 \right) + 2 - 3\alpha_{S2} \Bigg) \Bigg. + \frac{3}{2} h_B\alpha_{B2} = 0 \Rightarrow \\
               &\left(\frac{h_S^2 + h_Sl_S + l_S^2}{h_S + l_S)} \right) \Bigg( \Bigg. \frac{\alpha_{B1}\alpha_{S2}}{2\alpha_{B2}} + \frac{\alpha_{S1}}{2} + 1 - 3\alpha_{S2} \Bigg) \Bigg. + \frac{3}{2} h_B\alpha_{B2} = 0 \Rightarrow \\
               & \Bigg( \Bigg. 2 + \alpha_{S1} - 6\alpha_{S2} + \frac{\alpha_{B1}\alpha_{S2}}{\alpha_{B2}} \Bigg) \Bigg. \left( \frac{h_S^2 + h_Sl_S + l_S^2}{h_S + l_S} \right) + 3h_B\alpha_{B2} = 0 \numberthis \label{eq:case4_gen4}
           \end{align*}
           %
        
            % \begin{remark}
            %     If we perform the similar analysis using Equations \ref{eq:Ub_extended} and \ref{eq:Us_extended}, we get the following equations, 
            %     %
            %   \begin{align*}
            %       &\alpha_{S1}\alpha_{B2} + \alpha_{S2} (2 - 3\alpha_{B1}) = 0 \numberthis \label{eq:case4_gen1} \\
            %       &\Bigg( \Bigg. 2 + \alpha_{B1} - 6\alpha_{B2} +  \frac{\alpha_{S1}\alpha_{B2}}{\alpha_{S2}} \Bigg) \Bigg. \left( \frac{h_B^2 + h_Bl_B + l_B^2}{h_B + l_B} \right) + 3l_S\alpha_{S2} = 0 \numberthis \label{eq:case4_gen2} \\
            %       &\alpha_{B1}\alpha_{S2} + \alpha_{B2}(2 - 3\alpha_{S1}) = 0 \numberthis \label{eq:case4_gen3} \\
            %       & \Bigg( \Bigg. 2 + \alpha_{S1} - 6\alpha_{S2} + \frac{\alpha_{B1}\alpha_{S2}}{\alpha_{B2}} \Bigg) \Bigg. \left( \frac{h_S^2 + h_Sl_S + l_S^2}{h_S + l_S} \right) + 3h_B\alpha_{B2} = 0 \numberthis \label{eq:case4_gen4}
            %   \end{align*}
            %   %
            % \end{remark}
           
           \begin{remark}
               For a specific case of $\theta_B \sim U[0,1]$ and $\theta_S \sim U[0,1]$, replacing $l_B = l_S = 0$ and $h_S = h_B = 1$ in the above equations, we get the following set of equations,
               \begin{align*}
                   & \alpha_{B2}\alpha_{S1} + \alpha_{S2} (2 - 3\alpha_{B1}) = 0 \numberthis \label{eq:case4_1} \\
                   & \alpha_{S2}(2 + \alpha_{B1} - 6\alpha_{B2}) + \alpha_{S1}\alpha_{B2} = 0 \numberthis \label{eq:case4_2} \\
                   & \alpha_{B1}\alpha_{S2} + \alpha_{B2}(2 - 3\alpha_{S1}) = 0 \numberthis \label{eq:case4_3} \\
                   & \alpha_{B2}(2 + \alpha_{S1} - 6\alpha_{S2}) + \alpha_{B1}\alpha_{S2} + 3\alpha_{B2}^2 = 0 \numberthis \label{eq:case4_4}
               \end{align*}
               
               By solving these equations, we get $\alpha_{B1} \approx 0.882782$, $\alpha_{B2} \approx 0.588521$, $\alpha_{S1} \approx 1.2207$ and $\alpha_{S2} \approx 1.10806$.
               \end{remark}
           
           We summarise the above discussion as the following theorem,
           \begin{theorem} \label{main_result}
                 For a single-buyer single-seller two-unit $k$-double auction with $k$ = 0.5; $\theta_B \sim \mathbf{U} [l_B,h_B]$ and $\theta_S \sim \mathbf{U}[l_S,h_S]$, respectively; when they deploy scale based bidding strategies $b_B$ and $b_S$ and fix their scaling factors before even seeing their true types, below set of equations characterise the equilibrium,
                 \begin{itemize}
                     \item If $\alpha_{B1} = \alpha_{B2} = \alpha_B$ and $\alpha_{S1} = \alpha_{S2} = \alpha_S$, then Equation~\ref{eq:case1_alphab} and \ref{eq:case1_alphas} characterise BNE.
                     \item If $\alpha_{B1} \ne \alpha_{B2}$ and $\alpha_{S1} = \alpha_{S2} = \alpha_S$, then Equation~\ref{eq:case2_gen1}, \ref{eq:case2_gen2} and \ref{eq:case2_gen3} characterise BNE.
                     \item If $\alpha_{B1} = \alpha_{B2} = \alpha_B$ and $\alpha_{S1} \ne \alpha_{S2}$, then Equation~\ref{eq:case3_gen1}, \ref{eq:case3_gen2} and \ref{eq:case3_gen3} characterise BNE.
                     \item If $\alpha_{B1} \ne \alpha_{B2}$ and $\alpha_{S1} \ne \alpha_{S2}$, then Equation~\ref{eq:case4_gen1}, ~\ref{eq:case4_gen2}, ~\ref{eq:case4_gen3} and ~\ref{eq:case4_gen4} characterise BNE.
                 \end{itemize}
           \end{theorem}
           
           \begin{corollary}
                 For a single-buyer single-seller two-unit $k$-double auction with $k$ = 0.5; $\theta_B \sim \mathbf{U} [0,1]$ and $\theta_S \sim \mathbf{U}[0,1]$, respectively, when they deploy scale based bidding strategies $b_B$ and $b_S$ and fix their scaling factors before even seeing their true types, below scale factors constitute the BNE for each case,
                 \begin{itemize}
                     \item If $\alpha_{B1} = \alpha_{B2} = \alpha_B$ and $\alpha_{S1} = \alpha_{S2} = \alpha_S$, then $\alpha_B = \frac{2}{3}$ and $\alpha_S = 1$ constitute BNE.
                     \item If $\alpha_{B1} \ne \alpha_{B2}$ and $\alpha_{S1} = \alpha_{S2} = \alpha_S$, then $\alpha_{B1} = \frac{6}{7}$, $\alpha_{B2} = \frac{4}{7}$ and $\alpha_S = \frac{268}{315}$ constitute BNE.
                     \item If $\alpha_{B1} = \alpha_{B2} = \alpha_B$ and $\alpha_{S1} \ne \alpha_{S2}$, then $\alpha_B = \frac{2}{3}$, $\alpha_{S1} = 1$ and $\alpha_{S2} = 1$ constitute BNE.
                     \item If $\alpha_{B1} \ne \alpha_{B2}$ and $\alpha_{S1} \ne \alpha_{S2}$, then $\alpha_{B1} \approx 0.882782$, $\alpha_{B2} \approx 0.588521$, $\alpha_{S1} \approx 1.2207$ and $\alpha_{S2} \approx 1.10806$ constitute BNE.
                 \end{itemize}
           \end{corollary}

%%%%%%%%%%%%%%%%%%%%%%%%%%%%%%%%%%%%%%%%%%%%%%%%%%%%%%%%%%%%%%%%%%%%%%%%

\section{Building DDPG based Bidding Strategy}

    \hspace{0.5cm}In this section, we start with an overview of DDPG and then present arguments that motivated us to construct a bidding strategy using DDPG. We, then, describe \ddpg\  in detail along with the training and validation set-up for this novel bidding strategy. Additionally, we furnish experimental results that validate that DDPGBBS indeed converges to the theoretical equilibrium. Finally, we extend our \ddpg\ implementation intending to utilize it in smart-grids.
        
    %%%%%%%%%%%%%%%%%%%%%%%%%%%%%%%%%%%%%%%%%%%%%%%%%%%%%%%%%%%%%%%%%%%%%%%%

    \subsection{Deep Deterministic Policy Gradient (DDPG)}
    
\begin{figure}[ht]
          \centering
          \includegraphics[width=0.8\linewidth]{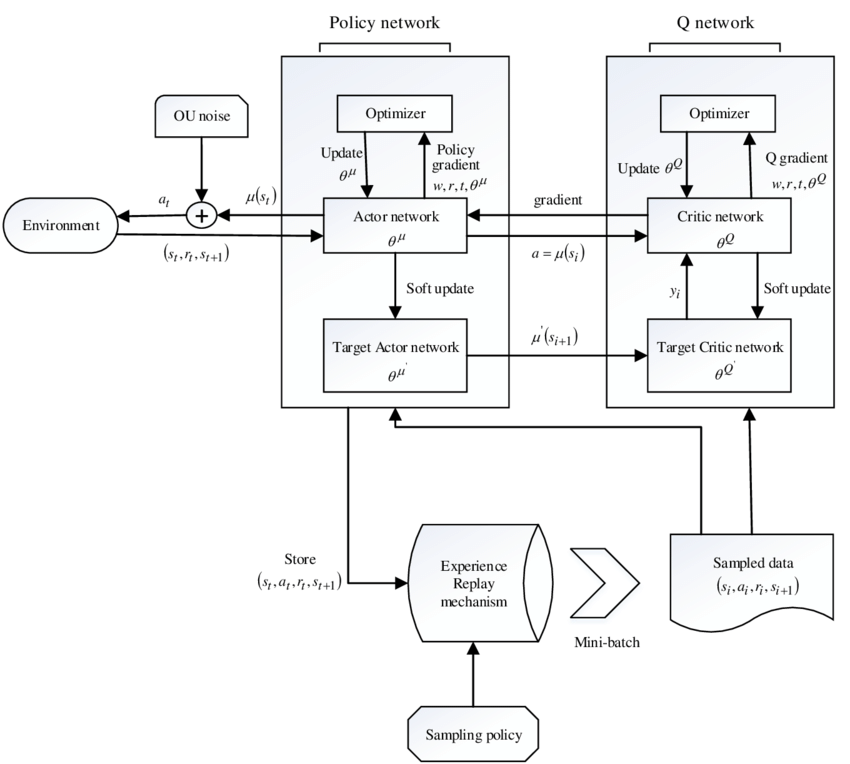}
          \caption{The DDPG Algorithm Structure Framework ~\cite{ddpgImage}}
          \label{fig:ddpg}
        \end{figure}

        \hspace{0.5cm}Deep Deterministic Policy Gradient ~\cite{lillicrap2016}, commonly known as DDPG, is an actor-critic, model-free reinforcement learning (RL) algorithm based on the concept of deterministic policy gradients ~\cite{SilverLHDWR14}. This algorithm works on continuous action spaces and has been proved to be very successful in a variety of continuous control problems ~\cite{lillicrap2016}, e.g., TORCS and MUJOCO environments. The algorithm deploys a parameterized policy (actor) and critic network that gets updated frequently using experiences gathered from sample roll-outs. While the policy network map learns an optimal deterministic mapping from state space to action space, the critic network learns the optimal value function using Bellman type updates used in Q-learning. Our choice of using the DDPG framework to solve the optimal bidding problem is due to our requirement that the bidding decisions of the proposed strategy are continuous in nature. Having motivated our choice of RL algorithm, we now proceed to describe the formulation of the bidding problem using the Markov decision process~\cite{puterman1994}. Figure~\ref{fig:ddpg} shows the structural framework of the DDPG algorithm.
        
    \subsection{DDPGBBS: A Bidding Strategy for Multi-unit Auctions}
    
        \hspace{0.5cm}In this section, we present details of our \ddpg\ based formulation. Here, the goal is to learn the buyer's NE bidding strategy, i.e., scale-factors of buyer for each of the possible scenarios discussed earlier. We propose the following MDP framework,
        
        \hspace{0.5cm}The state space $S$ consists of \emph{quantity to buy} $q$, where $q \in Q$, $Q = \{0, 1, 2\}$; and \emph{Buyer's true type} $\theta \in [0,1]$. The actions are the buyer's scale-factors  $\alpha_{B1}$, $\alpha_{B2} \in [0,1]$, which we want \ddpg\ to learn. Buyer receives reward $r = 0$ if no market-clearing happens, else it receives reward $r = -cp * cq$ (where $cp$ and $cq$ are clearing price and buyer's clearing quantity). As we are working with negative rewards, an optimal strategy would be one that maximizes the expected reward. A state $s$ transition to the next state where \emph{quantity to buy} is the remaining quantity $q'$ after auction-clearing in state $s$,  ($q' \in Q = \{0, 1, 2\}$), while \emph{buyer's true type} remains the same. Here we are considering single-shot auction, thus MDP terminates after a single iteration. Since, there is no possible action from the terminal state, a buyer will receive a terminal reward $r = -q' * \theta$.
        
        \hspace{0.5cm}\ddpg\ learns based on the clearing price information received from the clearing mechanism. We designed the reward function in such a way that if the buyer can get the items cleared in the auction, the buyer would pay lesser than its true value (lower clearing price); else, would buy the item at the buyer’s true value (higher clearing price). The task of \ddpg\ is to figure out the optimal scale factor to generate bids by looking at the variation in rewards for the selected scale factors; lower bids maximize the profit but with the risk of bids not getting cleared, and vice versa for higher bids. \ddpg\ learns which scale factors maximize the buyer's expected utility.
        
        \hspace{0.5cm}The \ddpg\ architecture details are as follows: the neural network used for actor-network had two hidden layers having 40 and 30 units, respectively. The critic networks also had two hidden layers with 40 and 30 units, respectively. Actions were not included until the second hidden layer of the critic network. Both networks used the \emph{ReLU} activation for all hidden layers. The actor-network used \emph{sigmoid}, while the critic network used the \emph{linear} activation in the output layer. The actor-network had two units in the input layer and two units in the output layer. We used \emph{Adam optimizer} for learning the neural network parameters and kept the learning rate of $10^{-4}$ and $10^{-3}$ for the actor and critic, respectively. we used a discount factor of $\gamma = 0.99$. For the soft target updates, we used $\tau = 0.001$.
    
    %%%%%%%%%%%%%%%%%%%%%%%%%%%%%%%%%%%%%%%%%%%%%%%%%%%%%%%%%%%%%%%%%%%%%%%%

    \subsection{Training and Validation Set-up of DDPGBBS}
    
        \hspace{0.5cm}In order to train our DDPGBBS, we designed a simulation of a multi-unit $k$-double auction, which includes the proposed scale-based bidding strategy of the buyer and the seller. Here, we aim to train DDPGBBS to learn the theoretical equilibrium of the buyer, but the same setup can be used to learn the seller's theoretical equilibrium as well. As shown in Section~\ref{analysis}, there are four possible scenarios based on scale factors of the buyer and seller. But, \textsf{Case-1} and \textsc{Case-3} lead to the same equilibrium; thus we consider them as a single scenario. 
        
        \hspace{0.5cm}Now, we train a separate \ddpg\  to learn the buyer's equilibrium for each of the three cases. We model the seller to play its scale-based equilibrium bidding strategy by fixing the scale factor according to theoretical results for each case, while its type is drawn from $U\sim[0,1]$. Then, we make our buyer select scale factors uniformly at random from the interval $[0,1]$ and select its type from $U\sim[0,1]$ and prepare bids using the scale-based bidding strategy. The simulated auction clearing mechanism clears the auction and decides the clearing price and quantity. After each auction clearing, the buyer receives a reward and information about the next state, and the episode terminates. A buyer prepares an experience quadruple $(s_t, a_t, r_t, s_{t+1})$ for each auction episode and stores it to a replay buffer. After sampling $r$ independent experience samples from replay buffer, we update the policy and critic networks. %we train \ddpg\  offline using the replay buffer. 
        We stop training after $n$ number of iterations; In our experiments we kept $r = 100000$ and $n = 10000$.
    
    %%%%%%%%%%%%%%%%%%%%%%%%%%%%%%%%%%%%%%%%%%%%%%%%%%%%%%%%%%%%%%%%%%%%%%%%

    \subsection{Validating Theoretical Results}
    
        \hspace{0.5cm}Here we discuss the results of the validation experiments. As discussed above, there are three possible scenarios and we trained a separate \ddpg\  model for each of the three cases. Table~\ref{tab:validation1} compares buyer's equilibrium scale factors learned by \ddpg\  against the theoretical scale factors for the case when $\alpha_{B1} = \alpha_{B2} = \alpha_B$ and $\alpha_{S1} = \alpha_{S2} = \alpha_S$. Similarly, Table~\ref{tab:validation2} compares buyer's equilibrium scale factors learned by \ddpg\  against the theoretical scale factors for the case when $\alpha_{B1} \neq \alpha_{B2}$ and $\alpha_{S1} = \alpha_{S2} = \alpha_S$. Finally, Table~\ref{tab:validation3} compares buyer's equilibrium scale factors learned by \ddpg\  against the theoretical scale factors for the general case when $\alpha_{B1} \neq \alpha_{B2}$ and $\alpha_{S1} \neq \alpha_{S2}$. Note that the buyer following \ddpg\  does not get to see seller's asks (neither during training nor during validation), and only learns the equilibrium strategy based on the feedback it receives from the auction clearing mechanism for its own actions. These results demonstrate that \ddpg\ is able to detect the choice of seller's scale factors and effectively learn the NE strategy for each case.
        
         \begin{table}[t]
          \centering
          \caption{Experimental scale factor values of buyer compared with theoretical scale factor values for case-1}
          \label{tab:validation1}
          \renewcommand{\arraystretch}{1.2}
          \begin{tabular}{||c | c | c||}\toprule
            \textit{} & \textit{} & \textit{$\alpha_{B1}$} \\ \midrule
            Theoretical Equilibrium & & 0.666667 \\
            \hline
            DDPG Empirical Equilibrium & mean & 0.577733 \\
            \hline
            & std & 0.023132 \\ \bottomrule
          \end{tabular}
        \end{table}
        
        \begin{table}[t]
          \centering
          \caption{Experimental scale factor values of buyer compared with theoretical scale factor values for case-2}
          \label{tab:validation2}
          \renewcommand{\arraystretch}{1.2}
          \begin{tabular}{||c | c | c | c||}\toprule
            \textit{} & \textit{} & \textit{$\alpha_{B1}$} & \textit{$\alpha_{B2}$} \\ \midrule
            Theoretical Equilibrium & & 0.857143 & 0.571428 \\
            \hline
            DDPG Empirical Equilibrium & mean & 0.928814 & 0.477797 \\
            \hline
            & std & 0.022228 & 0.046001 \\ \bottomrule
          \end{tabular}
        \end{table}
        
        \begin{table}[t]
          \centering
          \caption{Experimental scale factor values of buyer compared with theoretical scale factor values for case-3}
          \label{tab:validation3}
          \renewcommand{\arraystretch}{1.2}
          \begin{tabular}{||c | c | c | c||}\toprule
            \textit{} & \textit{} & \textit{$\alpha_{B1}$} & \textit{$\alpha_{B2}$} \\ \midrule
            Theoretical Equilibrium & & 0.882782 & 0.588521 \\
            \hline
            DDPG Empirical Equilibrium & mean & 0.855816 & 0.302352 \\
            \hline
            & std & 0.050594 & 0.221870 \\ \bottomrule
          \end{tabular}
        \end{table}
        
        \hspace{0.5cm}From Table~\ref{tab:validation1}, we can see that the empirical result we obtained using \ddpg\ is within $12.2\%$ of theoretical $\alpha_{B1}$. From Table~\ref{tab:validation2} (\ref{tab:validation3}), learned $\alpha_{B1}$ and $\alpha_{B2}$ are within $8.28\%(3.13\%)$ and $16.31\%(47.6\%)$ of the theoretical values, respectively. As we can see, except for $\alpha_{B2}$ in Table~\ref{tab:validation3}, DDPGBBS results are reasonably close to the theoretical results. As a comparison, the results presented by Susobhan et al. (in Table 1) \cite{susobhan20} for single-unit k-double auction showed $20.19\%$ and $24.72\%$ difference between empirical results and theoretical values for buyer and seller, respectively. Additionally, \ddpg\ showed low variances for all the scale factors except $\alpha_{B2}$ in Table~\ref{tab:validation3}, reinforcing our \ddpg’s stability.  
    
    %%%%%%%%%%%%%%%%%%%%%%%%%%%%%%%%%%%%%%%%%%%%%%%%%%%%%%%%%%%%%%%%%%%%%%%%

    \subsection{Extending DDPGBBS to Design Bidding Strategies for Smart-grids}
    
        \hspace{0.5cm} As PDA allows multiple auction instances for a delivery slot, our \ddpg\ needs to be updated accordingly; thus, we propose \emph{Extended \ddpg} to be helpful in general PDAs. Below is the updated MDP to use in PDA,  
        
        \hspace{0.5cm}The state space $S$ consists of \emph{proximity} $p$, where $p \in P$, $P = \{0, 1, 2, ..., 24\}$; \emph{quantity to buy} $q$, where $q \in R$; and \emph{Buyer's true type} $\theta \in R$. Here, we consider the buyer's true type as \emph{average unit balancing price for buying} from balancing market in a game. The actions are the buyer's scale-factors  $\alpha_{B1}$, $\alpha_{B2} \in [0,1]$. During the game, Extended \ddpg\ outputs these two scale-factors which gets multiplied with buyer's true valuation to form the two bids in the auction, while required bidding quantity is equally distributed into these two bids.
        
        \hspace{0.5cm}The reward depends on the current state $s$. If $s$ is not a terminal state, our buyer receives reward $r = 0$ if no market-clearing happens, else it receives reward $r = -cp * cq$ (where $cp$ and $cq$ are clearing price and buyer's clearing quantity); if $s$ is a terminal state then it receives reward $r = -q' * \theta$, where $q'$ is the remaining quantity at the end of all $24$ auctions. Again, as we are working with negative rewards, an optimal strategy would be the one that maximizes the expected reward. After each auction, state transition takes place, and MDP moves to the next state. \emph{Proximity} changes $p$ to $p-1$, \emph{quantity to buy} becomes the remaining quantity after current auction-clearing, and \emph{buyer's true type} remains the same. The episode ends when the buyer reaches the terminal state $T$, the buyer moves to $T$ either when \emph{Proximity} becomes zero or when the buyer buys all the required quantity.

        \hspace{0.5cm}We train this Extended \ddpg\ in offline fashion by collecting experiences in the replay buffer, using the PowerTAC PDA simulator. To collect experiences, we run two sets of experiments, with 20 games in each set. We make the Extended \ddpg\ to play against a ZI broker in two-player games in the first set. In the second set, Extended \ddpg\ competes against three other ZI brokers in four-player games. In each set, we distribute hourly demand equally between all the competing brokers; as a result, each broker had to participate equally in the wholesale market PDA. Extended \ddpg\  updates the replay buffer after each auction instance in the game. We trained our Extended \ddpg\ against ZI brokers as they do not follow any particular bidding pattern, and thus Extended \ddpg\ gets to see a wide range of states in the state-space $S$, which improves its learning. After the execution of both sets is completed, we update our Extended \ddpg\ using the combined replay buffer of both sets by following the standard DDPG update procedure used for validation experiments. 

%%%%%%%%%%%%%%%%%%%%%%%%%%%%%%%%%%%%%%%%%%%%%%%%%%%%%%%%%%%%%%%%%%%%%%%%

\section{Experiments and Results}

    \hspace{0.5cm}Here, we analyze the efficacy of Extended \ddpg\ in PDA by leveraging the PowerTAC PDA simulator. We first explain the experimental setup followed by the results of the offline experiments. To compare the performance of Extended \ddpg, we isolate the PowerTAC wholesale market while keeping the default market participants. We benchmark the performance of Extended \ddpg\ against the state-of-the-art and baseline strategies mentioned in Section \ref{ssec: powertac}. 

\begin{figure}[ht]
      \centering
      \includegraphics[width=0.45\linewidth]{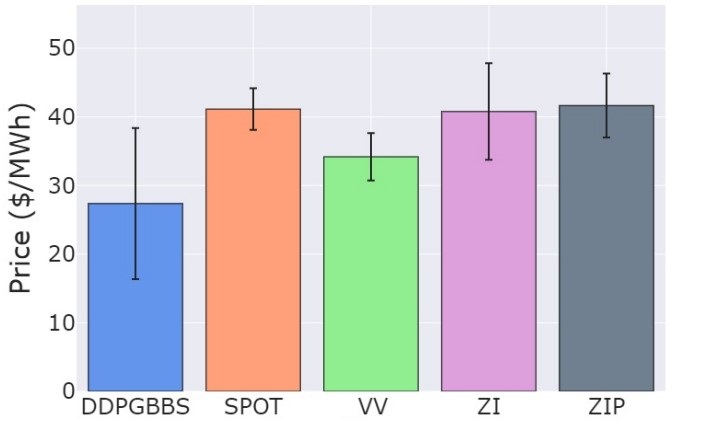}
      \caption{Average Unit Clearing Price Comparison for five-player game configuration}
      \label{fig:comparison}
    \end{figure}

    \hspace{0.5cm}We perform two batches of experiments; the first batch of experiments are divided into four sets. In each of these four sets, we play ten two-player games where Extended \ddpg\ is one of the two brokers in each set, while the second broker is selected from the list of four benchmarking brokers. In other words, Extended \ddpg\ competes against each benchmarking broker in a set of 10 two-player games. Similarly, in the second batch, we play ten five-player games having all the available brokers in the game. Table~\ref{tab:batch1} summarises the results of the first batch of experiments; it shows the normalized average unit clearing price of each set with respect to the Extended \ddpg's clearing price. A value greater than $1$ indicates that the competing broker in that set had a higher average clearing price than Extended \ddpg\, after playing ten games. Based on the results in Table~\ref{tab:batch1}, we observe that Extended \ddpg\ outperforms all the other bidding strategies consistently by at least $9.9\%$ in two-player games while achieving 33.76\% improvement against ZI. Figure~\ref{fig:comparison} shows the result for the second batch of experiments, where we compare the average unit clearing price of each broker across ten games in a five-player game configuration. The error bars on top of bar plots show the standard deviation of average clearing prices in 10 games. Here too, Extended \ddpg\  consistently outperforms all the other brokers by 21.42\% against second performing VV, while achieving almost 46\% improvement against other brokers. \ddpg\ obtained better results than the state-of-the-art strategies (VV and SPOT) used for comparison. \ddpg\ improves in the range $[21.42\%, 46\%]$ \ref{fig:comparison} over the baseline strategies as well as over VV and SPOT. Susobhan et al.'s VV (in Figure 1 \cite{susobhan20}) improved in the range $[-11.46\%, 22.01\%]$, while Chowdhury et al. (in Figure 3 in \cite{spot18}) improved in the range $[31.28\%, 32.7\%]$ over previous strategies.  It is also to be noted that, unlike some of the baseline brokers, Extended \ddpg\ does not incorporate any additional heuristics in its bidding strategy and still outperforms some of the best brokers of PowerTAC. 
    
    \begin{table}[t]
      \centering
      \caption{ Relative Unit Clearing Price Comparison}
      \label{tab:batch1}
      \renewcommand{\arraystretch}{1.2}
      \begin{tabular}{||c|c | c | c | c||}\toprule
        \textbf{\ddpg} & SPOT & VV & ZI & ZIP \\
        \hline
        \textbf{1.0} & 1.2832 & 1.0992 & 1.3376 & 1.1920 \\ \bottomrule
      \end{tabular}
    \end{table}

%%%%%%%%%%%%%%%%%%%%%%%%%%%%%%%%%%%%%%%%%%%%%%%%%%%%%%%%%%%%%%%%%%%%%%%%

\section{Conclusion}

    \hspace{0.5cm}In this paper, we first characterized the Bayesian Nash equilibrium for single-buyer single-seller two-unit single-shot k-Double auctions, where the buyer and the seller follow scale-based bidding strategies. We analyzed all possible cases resulting from how the buyer and the seller select their scale-factors in Theorem~\ref{main_result}. As the equilibrium analysis becomes intractable with the increase in the number of participating players and items, we presented a deep deterministic policy gradient bidding strategy \ddpg\ that is easily extendable to a real-world PDA. We experimentally validated that \ddpg\ achieves BNE approximately for each case. Furthermore, we proposed an extension of \ddpg\ that can be utilized in the real-world PDA. Finally, to examine the Extended \ddpg\ efficacy, we benchmarked it against several baselines and the state-of-the-art bidding strategy of PowerTAC wholesale PDA. We showed that it consistently outperforms some of the best PowerTAC wholesale bidding strategies.

%%%%%%%%%%%%%%%%%%%%%%%%%%%%%%%%%%%%%%%%%%%%%%%%%%%%%%%%%%%%%%%%%%%%%%%%

\bibliographystyle{unsrt}
\bibliography{main}

\end{document}